\documentclass{tdp}

\begin{document}

\title{The Privacy Policy Permission Model:
\newline
A Unified View of Privacy Policies}
\author{Maryam Majedi$^{*}$, Ken Barker$^{**}$}
\address
{
$^{*}$Department of Computer Science, University of Toronto, Toronto, ON, M5S 2E4, Canada. \\ 
$^{**}$Department of Computer Science, University of Calgary, Calgary, AB, T2N 1N4, Canada.\\
E-mail: {\small \tt{majedi@cs.toronto.edu}}, {\small \tt{kbarker@ucalgary.ca}}
}

%% This part here is not needed for submission
\TDPRunningAuthors{Maryam Majedi, Ken Barker}
\TDPRunningTitle{The Privacy Policy Permission Model: A Unified View of Privacy Policies}
\TDPThisVolume{1}
\TDPThisYear{2021}
\TDPFirstPageNumber{1}
\TDPSubmissionDates{Received 5 January 2021; received in revised form 22 April 2021; accepted 27 April 2021}

\maketitle

\begin{abstract}
Organizations use privacy policies to communicate their data collection practices to their clients. A privacy policy is a set of statements that specifies how an organization gathers, uses, discloses, and maintains a client's data. However, most privacy policies lack a clear, complete explanation of how data providers' information is used. We propose a modeling methodology, called the \textit{Privacy Policy Permission Model} (PPPM), that provides a uniform, easy-to-understand representation of privacy policies, which can accurately and clearly show how data is used within an organization's practice. Using this methodology, a privacy policy is captured as a diagram. The diagram is capable of highlighting inconsistencies and inaccuracies in the privacy policy. The methodology supports privacy officers in properly and clearly articulating an organization's privacy policy.
\end{abstract}


\section{Introduction}
We live in the age of data monitoring where our cell phones are personal tracking devices that record every movement, activity, and conversation. Social networks can capture a holistic representation of our lives, even if we are not direct participants. Babies may even have profiles on social networks before they are born. Corporations collect our activity information and permanently store it in their repositories. The colossal amount of gathered information is analyzed for predicting and often influencing our decisions. Modern technology, while promising a richer and easier way of life, provides a means of control of our information by large corporations such as Facebook\texttrademark, Amazon\texttrademark, Google\texttrademark, and Apple\texttrademark. The reality is that we are being tracked; but this does not always seem to unsettle us until the consequences are revealed through various forms of data breech or misuse. Most of us like our devices tailor-made to provide us with immediate, relevant information. We use wearable gadgets that measure our fitness, sleep, even happiness; or anything else that businesses envision as a value proposition for us. We often willingly trade our information for these services, not realizing how our data is used and analyzed, or considering the consequences. It is time to assess how our privacy has been compromised and what we can do to mitigate the negative consequences.

\section{Problem Definition}
Privacy policies are difficult to understand because they are typically long, vague, incorporate jargon, and do not clearly define how the data provider's information is used \cite{wu2010analysis},\cite{wilson2016creation}, \cite{oltramari2018privonto}. In addition, organizations might aggregate data to produce new (correct or incorrect) knowledge, about the data providers \cite{kim2002building}. Unfortunately, privacy policies are often ignored by users. The process of deciphering privacy policies is difficult, partly because they are written in natural language with all its inherent connotations, so the risk of misunderstanding is high. They may imply access permissions that are illegitimate, which could result in further privacy violations.

We introduce a modeling methodology that provides a standard representation of privacy policies that can accurately and clearly describe how data is used within an organization. Modeling privacy policies, and capturing all the syntactic and semantic aspects within their context, is a critical step to enforcing them.

Organizations use databases to store and manage data for their businesses. Traditional database design begins by developing a conceptual schema using a tool such as Entity Relationship Diagrams (ERD) \cite{chen1976entity}. Unfortunately, there are no tools that explicitly capture privacy, so a designer wishing to develop a privacy-preserving database must incorporate these in an \textit{ad hoc} way. Our approach is to develop a modeling methodology undertaken as a separate privacy-aware design step. This will explicitly incorporate privacy, independent from the database design, and facilitate the inclusion of privacy policies for all elements in the database.

\section{Contributions}
Motivated by traditional database design, (i.e. ERD), we develop and design a new privacy modeling methodology. Our approach is independent of the data domain so it can be used within any organization regardless of its activities. It produces a diagram that allows users to understand an organization's privacy policy. We demonstrate how a privacy policy is deconstructed into components and then implemented in a privacy diagram. Since this methodology produces diagrams that directly reflect an organization's privacy policies, privacy officers and administrators can use it to identify and resolve potential violations and contradictions. This modeling methodology is adaptable to policy changes by making the implications of alterations explicit in the resulting diagrams. 
Our contributions are as follows:
\begin{enumerate}

\item We develop a methodology called Privacy Policy Permission Model (PPPM) to depict privacy policies in a structured way including capturing:
\begin{enumerate}
\item Privacy components: 
    \begin{enumerate}
    \item[--]Roles: the categories within and beyond the organization that are granted access to private information. 
    \item[--]Purposes: the intentions behind the permitted accesses.
    \item[--]Data attributes: entities that are collected, accessed, and used.
    \end{enumerate}

\item Homogeneous connections:
    \begin{enumerate}
    \item[--]Role structure: indicates any hierarchy among roles.
    \item[--]Purpose structure: shows how purposes are constructed from ordered sets of tasks.
    \item[--]Attribute structure: indicates when attributes are aggregated and generate new information.
    \end{enumerate}
 
\item Heterogeneous connections (permissions)
    \begin{enumerate}
    \item[--]Role-purpose connections: indicate permissions for the roles to use purposes.
    \item[--]Purpose-attribute connections: capture connections of all purposes' tasks to their required data attributes.
    \end{enumerate}
\end{enumerate}

\item We describe how to model each of these components and how they are extracted and presented in a diagram using both a canonical example and a real-world example currently in use. 
\item We highlight the method's ability to identify gaps and contradictions by virtue of undertaking a gap analysis of the diagram produced.

\end{enumerate}
We begin by providing a brief review of closely related research. 

\section{Background}
Agrawal \textit{et al.} \cite{agrawal2002hippocratic} identify ten principles of privacy when describing their proposal for Hippocratic Databases based on the US Privacy Act of 1974 \cite{bushkin1976privacy}. These principles are purpose specification, consent, limited collection, limited use, limited disclosure, limited retention, accuracy, safety, openness, and compliance. Similar values are articulated in the European Union Data Protection approach \cite{EUdataregulations2018} which is followed by Canada \cite{DeCew2018}. Privacy legislation often requires that data collectors inform data providers about the privacy policies they practice. In some sectors explicit legislation/regulation may be in place such as: the US Health Insurance Portability and Accountability Act (HIPAA) of 1996 \cite{Act1996Health1996},  Canada’s Health Information Act (HIA) \cite{2018HealthHIA}, Canada’s Personal Information Protection Act (PIPA) \cite{2020PersonalAct}, or Canada’s Personal Information Protection and Electronic Documents Act (PIPEDA) \cite{2019ThePIPEDA}. While these acts may be encoded in an organization's privacy policy, which are developed, at least in part, to communicate their practices to their users \cite{jensen2005privacy}, the policies often do not clearly describe individuals' data usage, but rather focus on protecting the organization from legal consequences, rather than protecting the users themselves \cite{anton2003lack}. 

To avoid the risks associated with the misuse of personal data, some individuals provide false information or create several email accounts to protect their privacy \cite{pollach2007s}. Culnan and Armstrong suggest that organizations must address their clients' privacy concerns to earn their trust \cite{culnan1999information}. In fact, if clients trust a website, they are more likely to prefer it to purchase merchandise \cite{hong2010influence}. Therefore, it is important that organizations demonstrate their commitment to protecting their client's privacy \cite{olivero2004privacy}.

Earp \textit{et al.} \cite{earp2007privacy} compare different methods for presenting privacy policies to online health care customers. In their study, the users' perceptions of privacy policies are examined, and their understanding of categorized policies is measured. Finally, the results are compared to determine which representation is easier to understand. Their study shows that presenting the privacy policies in natural language is the most difficult to understand and is insufficient for conveying information to users. By categorizing the privacy statements, they increase the users' comprehension. Earp \textit{et al.} \cite{earp2007privacy} suggest that additional effort must be made to improve the way privacy policies are presented.

Often people find privacy policies too legalistic \cite{milne2004strategies}. Fabian \textit{et al.} \cite{fabian2017large} analyze the privacy policies of 50,000 popular websites and determine that the privacy policies are difficult to decipher. The notice-and-consent approach is widely used in the United States, but it is inadequate because it assumes that individuals read the privacy policies and understand the implications of providing data to the data collectors \cite{nissenbaum2011contextual}. 

VenkataSwamy \textit{et al.} \cite{venkataswamy2010cbpm} define data sets that are subject to the same policies, and maintain permissions using a matrix. Silva \textit{et al.} \cite{da2016improving} introduce a multilanguage approach called RSLingo4Privacy to improve privacy policies. In their work, statements are classified to create a logically consistent equivalent, which facilitates a visual representation. The classification is then extracted based on the terms used in the privacy policy. Our methodology models privacy policies independent of the data domain, according to the principle of data independence.

Modeling is an essential step for developing systems. In addition to providing visual presentations, models can be used to define how a system would behave in various situations. For this reason, they can also be used for predicting and understanding potential system gaps. Mai \textit{et al.} \cite{mai2018modeling} propose a modeling method to structure and analyze privacy and security specification requirements in the health care domain, which is useful when developing the software. Context is essential for understanding the privacy-protection being afforded by these models \cite{lachner2019context}. Our methodology incorporates context to constrain the way that data collectors use and transfer the information collected based on explicit statements in the privacy policy. Chen \cite{chen1976entity}, in his seminal contribution, proposes a diagrammatic technique to model entities and their relationships. This technique is independent of the entities' domains. Our approach is fundamentally inspired by this abstract, generic modeling tool.

\section{PPPM: An ERD for Privacy Policies}
\begin{figure}[H]
 \centering
	\includegraphics{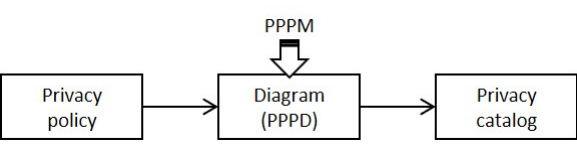}
	\caption{PPPM as a modeling methodology}
	\label{fig:PrivacyPolicy2PPPM2Catalog}
\end{figure}

This section introduces our methodology, Privacy Policy Permission Model, and describes how it can be used to create a unified and natural view of privacy policies. The model achieves a high degree of entity and domain independence. By providing a visual demonstration of privacy policies, our methodology highlights shortfalls to help organizations clearly explain their privacy policy statements. The resulting Privacy Policy Permission Diagram (PPPD) can then be implemented in a privacy catalog (see Figure \ref{fig:PrivacyPolicy2PPPM2Catalog}). The privacy catalog, our current project, is a set of database tables that capture a modeled privacy policy. Once the diagram is generated, it is used to populate the privacy catalog in the database. This catalog can then be used for granting or revoking accesses to the data layer, thereby enforcing privacy policies. We will introduce it in future work.
\newline
\clearpage
\subsection{Privacy Policy Components and Connections}
A privacy policy consists of policy statements. These statements are natural language sentences that define how the data in an organization's custody must be maintained and used. Each privacy statement consists of privacy components. Privacy components are categorized differently by different researchers. Barker \textit{et al.} \cite{barker2009data} introduce a data privacy taxonomy to describe key privacy components. According to their taxonomy, the privacy components are purpose, visibility, granularity, and retention. Ni \textit{et al.} \cite{ni2007conditional}, \cite{ni2008obligation} consider obligation and condition as privacy policy components as well. We further consider a data attribute a privacy component. In this work, we refer to visibilities as roles. We also develop granularity as functions that apply to data items to prepare them for purposes' usage, and we develop retention as conditions for accessing data. Thus, we do not consider granularity or retention as separate privacy components. Moreover, this methodology does not support obligations at this time. Therefore, the privacy components in this work are identified as roles, purposes, and data attributes.

We now formally describe each component (Section \ref{sec:Components}) and explain how connections between homogeneous components are mapped in a \textit{Privacy Policy Permission Diagram} (PPPD) (Section \ref{sec:HomogeneousConnections}). Finally, we show how the heterogeneous connections, and any conditions associated with them, are added to the diagram (Section \ref{sec:HeterogeneousConnections}).

\subsubsection{Components}\label{sec:Components}

\paragraph{Roles}\mbox{}

\textit{Roles} define the category of subjects (i.e., individuals, organizations, or agents) who are accessing data. Roles are identified as visibility in the taxonomy developed by Barker \textit{et al.} \cite{barker2009data}. An organization's functionality defines the appropriate visibility categories. In this work, a role is an application-specific actor that is identified in the written policy statements, and is a subject accessing data. We denote the set of roles: {$R = \{r_1, r_2, \ldots, r_n \}$}, where $r_i$ denotes a specific role identified in the policy statement. 

\paragraph{Purposes}\mbox{}

\textit{Purposes} are reasons to access data. For example, a bank clerk's purpose for accessing a customer's account information might be 'Issuing a statement'. In other words, purposes are reasons for data collectors to gather and use data from providers. We define the set of all purposes as: {$P = \{p_1, p_2, \ldots, p_n \}$}, where $p_i$ denotes a specific purpose identified in the policy statement. 

\paragraph{Data Attributes}\mbox{}

\textit{Data attributes} represent specific pieces of sensitive information. A \textit{data item} is an instance of a data attribute. For example, 'Age' is a data attribute, while '26' may be the value of a data item. A data attribute's sensitivity level is relative depending on the data provider's concern/opinion about that data item's value. A data attribute is drawn from the written privacy policies and is an application-specific object used in an environment. For the sake of simplicity, we will use the term 'attribute' rather than the more precise term 'data attribute' throughout this paper.

In an ERD, attributes are an entity's properties, and their values are set when they are instantiated. In a PPPD, there can be an arbitrary number of application-specific attributes ($d_i$), identified and collected from a policy statement into a set as {$D = \{d_1, d_2,\ldots, d_n \}$}.

A privacy policy might describe different categories for attributes which we call \textit{attribute groups}. These groups are used to refer to a set of attributes when specifying access permissions in a privacy policy. We denote attribute groups as: {$G = \{g_1, g_2, \ldots, g_n \}$}, where each $g_i$ denotes an attribute group defined in the privacy policy. Note that each attribute could belong to zero or more groups.

A data item's specificity can be modified by its granularity level. \textit{Granularity} specifies the precision used/needed when data is accessed for a purpose. For example, Barker \textit{et al.} \cite{barker2009data} categorize granularity as 'None', 'Existential', 'Partial', and 'Specific'. Granularity can be used to provide enhanced privacy by generalizing or making the data value more abstract. For example, granularity specifies whether 'age' should be provided as the exact age, or as an age range, such as 'Child', 'Teenager', or 'Adult'. Granularity has been addressed in both a deterministic and analytical way in the literature \cite{venkataramanan2016data} and our contributions will work in either methodology because it is undertaken at the policy level.In our model, granularity is supported by defining conversion functions that alter data precision based on the purpose for which it is accessed (see Purpose structure in the next Section). 

\subsubsection{Homogeneous Connections}\label{sec:HomogeneousConnections}
Roles, purposes, and attributes are first modeled independent of their interrelationships so we defined the connections within instances of the same type of components as homogeneous connections. The resulting structures help us clarify how instances of the same component type relate to one another. Each is described next.

\paragraph{Role Structure}\mbox{}

The role structure ($RS$) is based on the connections between roles. $RS$ is a set of ordered pairs, $r_i$,$r_j \in R$, when $r_i$ is superior\footnote{Superior in this context describes a reports-to relationship such as a manager-employee or a teacher-student model.} to $r_j$ and is depicted as: $r_i \rightarrow r_j$. Thus: $RS = \{(r_i, r_j) \ | \ r_i \rightarrow r_j\}$ for all $r_i$ and $r_j$ identified in the privacy policy that are connected. This relationship implies that $r_i$ holds at least all the access permitted to $r_j$. When $RS=\emptyset$ the roles are mutually exclusive with no explicit hierarchical structure. 

\paragraph{Purpose Structure}\mbox{}

Privacy concerns primarily arise due to access to data. Every such access should be based on its purpose. Purposes can access data in PPPM through \textit{tasks}. Thus, a task describes how a specific attribute value is used for a purpose. Therefore, we designate a task for every data usage, and then compose tasks into sequences that reflect the purpose's data accesses. Tasks can be composed in different ways for different purposes as illustrated in the following.

Thus, a purpose $p_i$ is an ordered set of tasks denoted as $p_i = (t_1, t_2, \ldots, t_n)$ where for each task $t_i \in T$ (where $T$ is the set of all tasks) exactly one attribute $d_j$ is accessed. Note however that a $d_j$ could be accessed by multiple tasks. 
Furthermore, each task is associated with a granularity function that modifies the attribute's value to match the required granularity level for the purpose's use. This allows us to capture purposes within context, and ultimately clarifies how attributes are used by each purpose.
\newline

\paragraph{Attribute Structure}\mbox{}

The attribute instances' structure represents how attributes are combined in an environment. If a privacy policy indicates that specific attributes are combined to generate new information, we establish a connection between them, and add the new information as an attribute. The attributes and their connections form a structure.
The attribute structure $(DS)$ is created based on the attributes that are aggregated to generate new information. $DS$ is a set of ordered triples where $d_i$, $d_j$, $d_k \in D$ where $d_i$ and $d_j$ are aggregated to generate a new data $d_k$. This aggregation is depicted as: $(d_i,d_j) \rightarrow d_k$. Thus: $DS=\{(d_i,d_j,d_k )\ |\ (d_i,d_j)\rightarrow d_k\}$ for all $d_i$ and $d_j$ identified in the privacy policy that aggregate to derive $d_k$ . When $DS=\emptyset$, no data attributes are aggregated in the policy. 

\subsubsection{Heterogeneous Connections (Permissions)}\label{sec:HeterogeneousConnections}
\textit{Heterogeneous connections} occur between components of different types. They either provide permissions when roles use purposes, or when purposes access data. Thus, we have: \textit{Role-purpose permissions} and \textit{Purpose-attribute permissions}. Permissions may have conditions, described in the following. 

\paragraph{Role-Purpose Permissions}\mbox{}

Role-purpose permissions provide permissions for roles to use purposes. If a role's access is conditional, its condition is added to the connecting role-purpose instance permission. Since this permission is effectively applied only to the connection between the role and purpose layers, the condition is independent of any attribute accessed.

We denote role-purpose permission $(RP)$ as a set of triples corresponding to a role $(r_i \in R)$, purpose $(p_j \in P)$, and condition $(c_k \in C)$ where $c_k$ may be null. Thus:

$RP=\{(r_i,p_j,c_k )\ |\ r_i$ \textit{and} $p_j$ \textit{is permitted under condition $c_k\}$}
as stated in the privacy policy.

\paragraph{Purpose-Attribute Permissions}\mbox{}

As described in Section \ref{sec:HomogeneousConnections}, homogeneous connections in the purpose layer create a structure based on purposes' ordered sets of tasks, such that each task requires exactly one attribute. In that stage, tasks were not connected to their attributes because tasks and attributes belong to different, heterogeneous layers. The purpose-attribute permissions in this stage connect the tasks to their required attributes, which capture the heterogeneous connections between the purpose and attribute layers.

Purpose-attribute permissions $(PD)$ can include a condition, which is placed on the corresponding task and attribute connection for that purpose. These conditions are independent of roles. For example, a purpose-attribute condition defining \textit{retention}, could be specified to expire data independent of a role access permission. Conditions must be satisfied in advance of access. This choice effectively eliminates using post-access obligations \cite{ni2008obligation} that are, in general, unenforceable, so this decision provides a stronger privacy guarantee. We leave for future research the incorporation of obligations once techniques are developed to enforce them.

Purpose-attribute permission $(PD)$ is a set of triples corresponding to a purpose $(p_i \in P)$, task $(t_j \in T)$, and condition $(c_k \in C)$ where $c_k$ may be \textit{null}. Thus:

$PD=\{(p_i,t_j,c_k )\ | \ p_i$ \textit{and} $t_j$ \textit{is permitted under condition} $c_k \}$ as stated in the privacy policy.
\newline

\paragraph{Conditions}\mbox{}

A \textit{condition} is a logical statement that must be satisfied to enact the purpose of a privacy policy. For example, Facebook's\texttrademark \ privacy policy requires that users be 13 years or older for registration. Thus, '$Age\ \geq \ 13$' is a conditional statement for 'Registration'. We denote the set of conditions as: {$C = \{c_1, c_2, ..., c_n\}$}.

\subsection{Extracting Information from the Privacy Policy}\label{sec:ExtractingInformation}
We illustrate the process of extracting privacy policy information using an imaginary organization that sells products online. The organization collects client information and ships orders. It analyzes client order lists and uses their age information to send birthday gifts. The complete privacy policy is provided in Appendix \ref{apx:ImgOnlShopPP}. For the purpose of illustration, we assume this privacy policy is complete, precise, and unambiguous.
Some privacy component instances and their connections are directly extracted from the statements. However, one must often infer them from the context. We describe three stages for capturing information from the written privacy policies. In the first stage, we capture components from the statements. The second stage captures homogeneous connections. The final stage defines heterogeneous connections or permissions.

In the following section, we describe each stage using our canonical privacy policy.

\subsubsection{First Stage: Capturing Component Instances}
This stage identifies all privacy component instances, and places them into their corresponding layer. In this section, we provide an example instance of each component type, and then identify all other instances occurring in the canonical policy.

\paragraph{Capturing Role Instances}\mbox{}

All role instances are captured and placed in the role layer. Example \ref{exm:StatementRole} illustrates a statement from \ref{sec:OrderShipment} of our online shopping privacy policy containing a role. 'Deliverer', in this statement, is the subject that accesses data to complete an action, so it is a role. 
\newline
\begin{example}
	\label{exm:StatementRole}
	Role as a subject.
	\begin{figure}[H]
		\raggedright
		\includegraphics{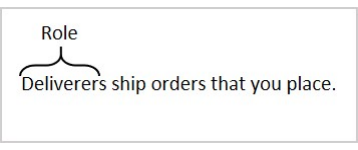}
	\end{figure}
\end{example}

All the roles identified in the privacy policy are recorded in Table \ref{table:RoleInstances}.
\begin{table}[H]
    \centering
    \scalebox{0.85}
    {
    \begin{tabular}{c l}
        \toprule
        \textbf{Label} & \textbf{Role}\\
        \midrule
            r\textsubscript1 & Manager \\
            r\textsubscript2 & Deliverer \\
            r\textsubscript3 & Analyzer \\
            r\textsubscript4 & Marketer \\
        \bottomrule 
    \end{tabular} 
    }
    \caption{Role instances} 
    \label{table:RoleInstances} 
\end{table}

\paragraph{Capturing Purpose Instances}\mbox{}

Example \ref{exm:StatementPurpose} illustrates the same policy statement in the articulation of the \textit{Shipment} purpose in our sample privacy policy. In this example, 'Shipment' is specified as the reason for accessing customer information by the deliverer. 
\newline
\begin{example}
	\label{exm:StatementPurpose}
	Purpose as an action.
	\begin{figure}[H]
		\raggedright
		\includegraphics{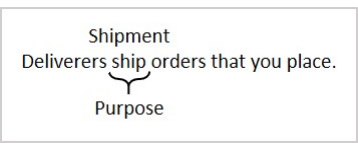}
	\end{figure}
\end{example}

Other purposes identified in this privacy policy (found in \ref{sec:AccessYourData}, \ref{sec:OrderShipment}, and \ref{sec:Marketing}) are shown in Table \ref{table:PurposeInstances}.
\newline

\begin{table}[H]
    \centering
    \scalebox{0.85}
    {
    \begin{tabular}{c l}
        \toprule
        \textbf{Label} & \textbf{Purpose}\\
        \midrule
            p\textsubscript1 & Shipment \\
            p\textsubscript2 & Marketing \\
            p\textsubscript3 & Sending gift \\
            p\textsubscript4 & Analyzing \\
        \bottomrule 
    \end{tabular} 
    }
    \caption{Purpose instances} 
    \label{table:PurposeInstances} 
\end{table}

\paragraph{Capturing Attribute Instances}\mbox{}

Example \ref{exm:StatementAttribute} shows a privacy policy statement from \ref{sec:InfoCollection} of the privacy policy that contains three attributes, 'Name', 'Email', and 'Address'. From this statement, we model an attribute group, 'Personal information', which includes all these attributes. 
\newline
\begin{example}
	\label{exm:StatementAttribute}
	Inferring attributes from a policy statement.
	\begin{figure}[H]
		\raggedright
		\includegraphics{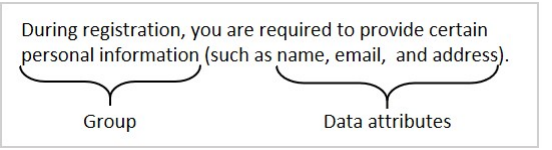}
	\end{figure}
\end{example}

'DOB' (Date of birth), 'Order list', 'Credit card information', and 'Interest' are other attributes mentioned in \ref{sec:OrderShipment} and \ref{sec:Marketing} of the privacy policy. 'Credit card information' also belongs to the 'Personal information' group but other attributes do not belong to any group. This information is recorded in Table \ref{table:AttributeInstances}.
\newline

\begin{table}[H]
    \centering
    \scalebox{0.85}
    {
    \begin{tabular}{c l l}
        \toprule
        \textbf{Label} &\textbf{Attribute} & \textbf{Group}\\
        \midrule
            d\textsubscript1 & Name & Personal information \\
            d\textsubscript2 & Order list & None \\
            d\textsubscript3 & Credit Card information & Personal information \\
            d\textsubscript4 & Address & Personal information \\
            d\textsubscript5 & Email &	Personal information \\
            d\textsubscript6 & DOB & Personal information \\
            d\textsubscript7 & Interest & None \\
        \bottomrule 
    \end{tabular} 
    }
    \caption{Attribute instances} 
    \label{table:AttributeInstances} 
\end{table}

\subsubsection{Second stage: Capturing Homogeneous Connections}
In this stage the component's layer structure is defined by capturing the connections between homogeneous component instances.

\paragraph{Capturing Role Layer Connections}\mbox{}

\ref{sec:Marketing} in the example privacy policy states that "The manager supervises analyzers and deliverers". This statement indicates that the 'Manager' role is superior to the 'Deliverer' and 'Analyzer' roles. Table \ref{table:RoleStructureInstances} shows all the relationships/connections between these roles in our sample privacy policy. 
\newline

\begin{table}[H]
    \centering
    \scalebox{0.85}
    {
    \begin{tabular}{l l}
        \toprule
        \textbf{Superior} & \textbf{Inferior}\\
        \midrule
            Manager & Deliverer \\
            Manager & Analyzer \\
            Analyzer & Marketer \\
        \bottomrule 
    \end{tabular} 
    }
    \caption{Roles connections} 
    \label{table:RoleStructureInstances} 
\end{table}

\paragraph{Capturing Purpose Layer Connections}\mbox{}

The policy statement in \ref{sec:OrderShipment} "To ship your orders, deliverers access your name, order list, credit card information, address, and email address to respectively identify you, process your order, charge fees, ship the parcel, and finally, inform you about the shipment", fully defines the tasks associated with the 'Shipment' purpose. Table \ref{table:PurposeStructureInstances} contains all the purpose layer connections to their tasks for the example privacy policy. 
\newline

\begin{table}[H]
    \centering
    \scalebox{0.85}
    {
    \begin{tabular}{l l}
        \toprule
        \textbf{Purpose} & \textbf{Tasks}\\
        \midrule
            Shipment &	Identify client, Process order list, Charge fees, Ship parcel, Inform client \\
            Analyzing &	Analyze based on age, Analyze shopping habit, Determine interest \\
            Marketing &	Identify client (Age \textgreater 18), Send advertisement \\
            Sending gift & 	Check DOB, Identify client, Ship parcel \\
        \bottomrule 
    \end{tabular} 
    }
    \caption{Purpose connections} 
    \label{table:PurposeStructureInstances} 
\end{table}

Task ordering is vital in this process because it describes purposes' proper data access order during execution.

\paragraph{Capturing Attribute Layer Connections} \mbox{}

Attribute connections are established if the policy specifies that they are combined. The resulting connected attributes create a structure. Attribute combinations often generate new information that an organization might use for its own purposes. For example, in \ref{sec:Marketing}, the policy states: "Our analyzers combine your date of birth, and shopping history to better understand your shopping habits, and predict your interests"\footnote{A customer's 'order history' is defined as the combination of previous 'order lists'.}, so 'DOB' and 'Order list' are combined and analyzed to predict customer 'Interest'. Table \ref{table:AttributeStructureInstances} illustrates combined attributes from the example policy. 
\newline

\begin{table}[H]
    \centering
    \scalebox{0.85}
    {
    \begin{tabular}{l l l}
        \toprule
        \textbf{Attribute 1} & \textbf{Attribute 2} & \textbf{New Attribute}\\
        \midrule
            DOB	& Order list &	Interest \\
        \bottomrule 
    \end{tabular} 
    }
    \caption{Attribute connections} 
    \label{table:AttributeStructureInstances} 
\end{table}

\subsubsection{Third stage: Capturing Heterogeneous Connections (Permissions)}
Connections between roles and purposes, and the permissions required for purposes to access attributes are called heterogeneous connections. We now describe how to capture these permissions and their conditions from the sample privacy policy. 
\paragraph{Capturing Role-purpose Permissions}\mbox{}

Role-purpose permissions must use statements indicating the purpose that a role has for access. These statements define permission connections between roles and purposes. In our example privacy policy, the statement "Deliverers ship orders that you place", specifies that there is a connection between the role 'Deliverer' and the purpose 'Shipment'. These statements may also include conditions. For example, the statement "Marketing staff members will send you advertisements within business hours," places a condition on the permission for the 'Marketer' role when using the purpose 'Marketing'. The condition specifies that the role 'Marketer' can perform 'Marketing' only between 8 am and 5 pm. This condition is independent of the attributes utilized.

Table \ref{table:RolePurposePermissionInstances} illustrates the completed role-purpose permissions list from our example. The privacy policy also indicates that an 'Analyzer' is allowed to use the purpose 'Analyzing', and that the 'Marketer' can perform the 'Sending gift' purpose.
\newline

\begin{table}[H]
    \centering
    \scalebox{0.85}
    {
    \begin{tabular}{l l l}
        \toprule
       \textbf{ Role} & \textbf{Purpose}& \textbf{Condition}\\
        \midrule
            Deliverer &	Shipment & \\	
            Analyzer & Analyzing & \\
            Marketer & Marketing & 8 am \textless now() \textless 5pm
 \\
            Marketer & Sending gift &	\\
        \bottomrule 
    \end{tabular} 
    }
    \caption{Role-purpose permissions} 
    \label{table:RolePurposePermissionInstances} 
\end{table}

\clearpage
\paragraph{Capturing Purpose-attribute Permissions}\mbox{}

Statements that specify a purpose using an attribute define purpose-attribute permissions. These permissions are captured by adding connections between the purpose's tasks and corresponding attributes, which may include conditions. Consider the statement in \ref{sec:Marketing} that "[a] customer's name can be used for marketing if the customer is over 18 years old." This statement indicates permission for the purpose 'Marketing' to allow access to the 'Name' attribute. The 'Marketing' purpose includes tasks 'Identify client' and 'Send advertisement' according to the policy. This infers that the tasks 'Identify client' must access the 'Name' attribute. In the next step, we place the 'Age \textgreater 18' condition on the purpose-attribute permission where the task 'Identify client' accesses 'Name'.

Other permissions are listed in Table \ref{table:PurposeAttributePermissionInstances}, where the first column contains the tasks, the second contains the data attributes that the tasks require, the third contains any conditions, and the last column contains the granularity function used to modify the data attribute for the tasks' usage. For example, the task 'Analyze based on Age' requires the age calculated from the attribute 'DOB'. 
\newline

\begin{table}[H]
    \centering
    \scalebox{0.85}
    {
    \begin{tabular}{l l l l l}
        \toprule
       \textbf{Label} & \textbf{Task} & \textbf{Attribute} & \textbf{Condition} & \textbf{Granularity}\\
        \midrule
            t\textsubscript1 & Identify client	& Name & Age \textgreater 18 \\
            t\textsubscript2 & Process order list & Order list & \\
            t\textsubscript3 & Charge fees & CC Info & \\
            t\textsubscript4 & Ship parcel & Address & \\
            t\textsubscript5 & Inform client & Email & \\
            t\textsubscript6 & Send advertisements & Email & \\
            t\textsubscript7 & Check DOB & DOB	& \\
            t\textsubscript8 & Analyze based on Age & DOB & & Date2Age \\ 
            t\textsubscript9 & Analyze shopping habit & Order list & \\	
            t\textsubscript{10} & Determine interest & Interest & \\
        \bottomrule 
    \end{tabular} 
    }
    \caption{Purpose-attributes permissions} 
    \label{table:PurposeAttributePermissionInstances} 
\end{table}

\clearpage
\subsection{Developing the Diagram}
We now develop the diagram for the example privacy policy using the extracted information from Section \ref{sec:ExtractingInformation}. We first create the three component layers and their homogeneous connections, and then complete the diagram by adding the permissions across heterogeneous component layers.

\subsubsection{Role Layer Diagram}
\begin{figure}[H]
 \centering
	\includegraphics[scale=0.90]{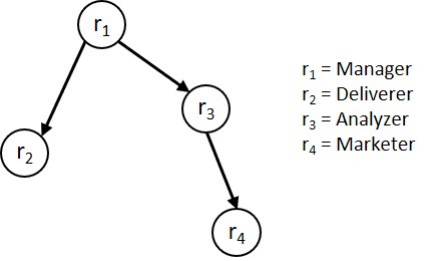}
	\caption{Role layer diagram}
	\label{fig:DiagramRole}
\end{figure}

Recall that Table \ref{table:RoleInstances} contains the four roles required to create the role layer diagram, whose connections, defined in Table \ref{table:RoleStructureInstances} , are captured in Figure \ref{fig:DiagramRole}. Directed edges from superior roles to inferior ones must be defined to complete the role layer diagram. To simplify this diagram, a node label is defined for each role and a legend is provided, listing the corresponding roles' instances. For example, in Figure \ref{fig:DiagramRole} label $r_1$ represents the role of 'Manager'. This practice is used in all the figures presented in the balance of the paper.
\newline

\subsubsection{Purpose Layer Diagram}
\begin{figure}[H]
 \centering
	\includegraphics[width=0.90\textwidth]{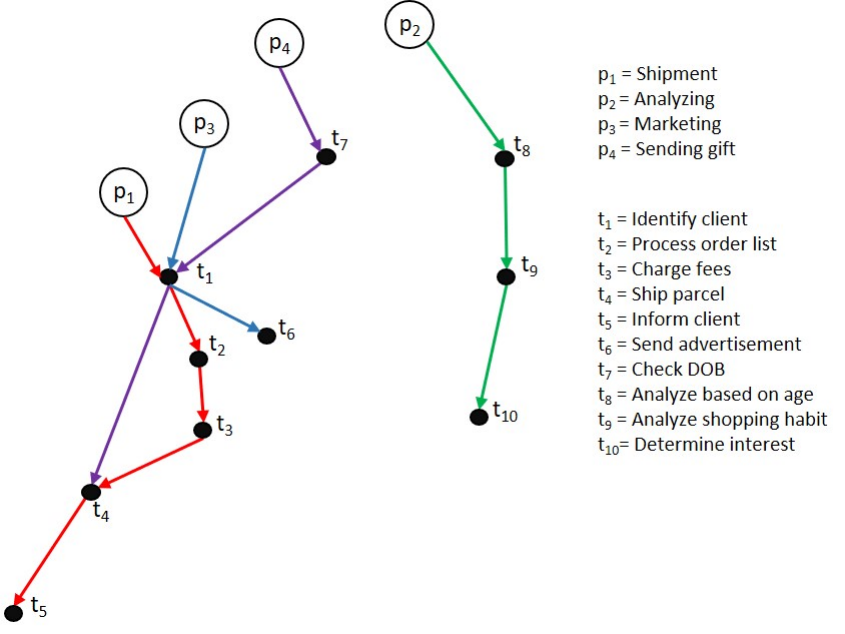}
	\caption{Purpose layer diagram}
	\label{fig:DiagramPurpose}
\end{figure}

The purpose layer contains purposes, represented as nodes (see Figure \ref{fig:DiagramPurpose}), which are connected to a sequence of tasks defined by the attributes they access. When a purpose's tasks (represented by solid dots) are specified in the privacy policy, they connect to their purpose through directed edges. If no tasks are specified for a particular purpose, only the purpose itself is included. The directed edges between task nodes capture the task order for a particular purpose. Recall that Table \ref{table:PurposeStructureInstances} identifies the purposes, along with their tasks and their orders. Figure \ref{fig:DiagramPurpose} uses '$p$'s to represent purposes, which are composed of '$t$'s, which represent tasks. For clarity, we use colour to illustrate different task sequences. 
\newline

\subsubsection{Attribute Layer Diagram}
\begin{figure}[H]
 \centering
	\includegraphics[width=0.60\textwidth]{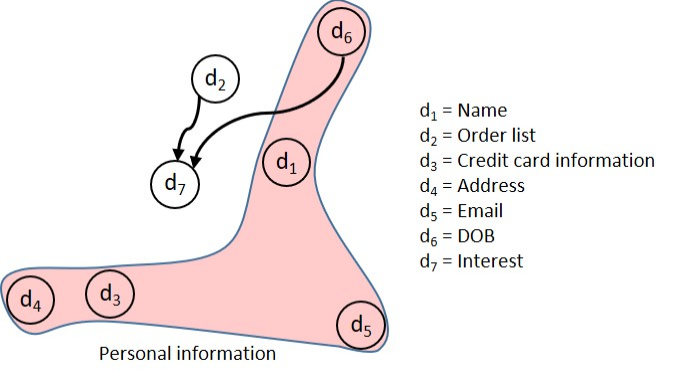}
	\caption{Attribute layer diagram}
	\label{fig:DiagramAttribute}
\end{figure}

In the attribute layer, attribute instances are nodes and their connections are edges. When attribute groups are specified, for the purpose of illustration, they are represented by surrounded areas containing their members. Recall that Table \ref{table:AttributeInstances} identifies the attributes in the policy. The 'Personal information' group, containing its attributes, is illustrated as a surrounded area in Figure \ref{fig:DiagramAttribute}.
\newline
Attribute layer connections capture statements in the privacy policy that indicate that the organization combines attributes in some way, which may give rise to implied attributes. From Table \ref{table:AttributeStructureInstances}, the 'Order list' and 'DOB' are combined, which gives rise to another attribute 'Interest'. Thus, the 'Interest' attribute is added to the diagram as a node and the arrows depict the two attributes that their combination produced. Figure \ref{fig:DiagramAttribute} illustrates the attribute layer which contains this combination.

\subsubsection{Role-purpose Permissions Diagram}
\begin{figure}[H]
 \centering
	\includegraphics[width=\textwidth]{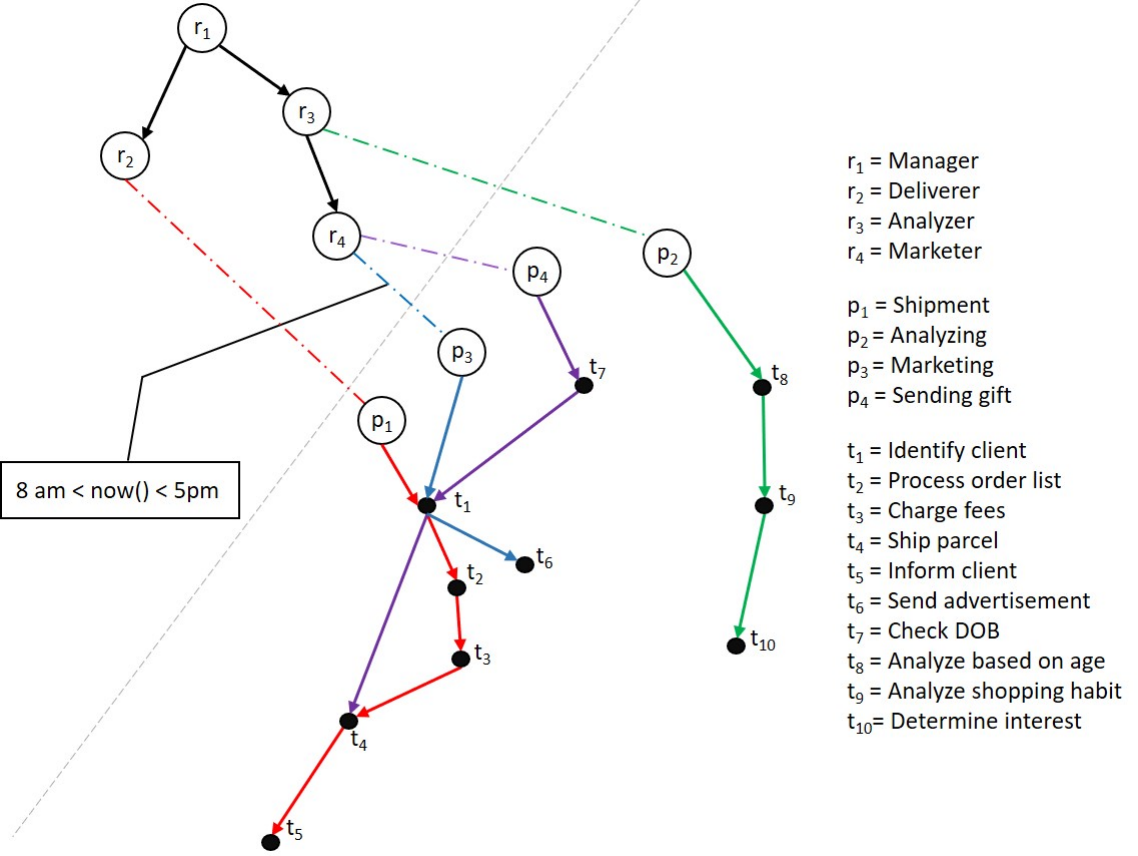}
	\caption{Role-purpose diagram}
	\label{fig:DiagramRolePurpose}
\end{figure}

Role-purpose heterogeneous connections and any corresponding conditions must be added to the diagram. Table \ref{table:RolePurposePermissionInstances} identifies these connections, which are illustrated in Figure \ref{fig:DiagramRolePurpose} with dashed-lines. For example, the 'Deliverer' role and the 'Shipment' purpose are connected, thereby illustrating that a deliver has legitimate permission to undertake shipment. Similarly, the 'Analyzer' role has permission to undertake the 'Analyzing' purpose; and the 'Marketer' role is connected to both the 'Marketing' and the 'Sending gift' purposes. To illustrate the use of conditions, the permission for the 'Marketer' role is conditional in that it must occur during the workday. These conditions are attached to the edges between roles and purposes as illustrated between $r_4$ and $p_3$ in Figure \ref{fig:DiagramRolePurpose}.
\newline

\subsubsection{Purpose-attribute Permissions Diagram}
\begin{figure}[H]
 \centering
	\includegraphics[width=\textwidth]{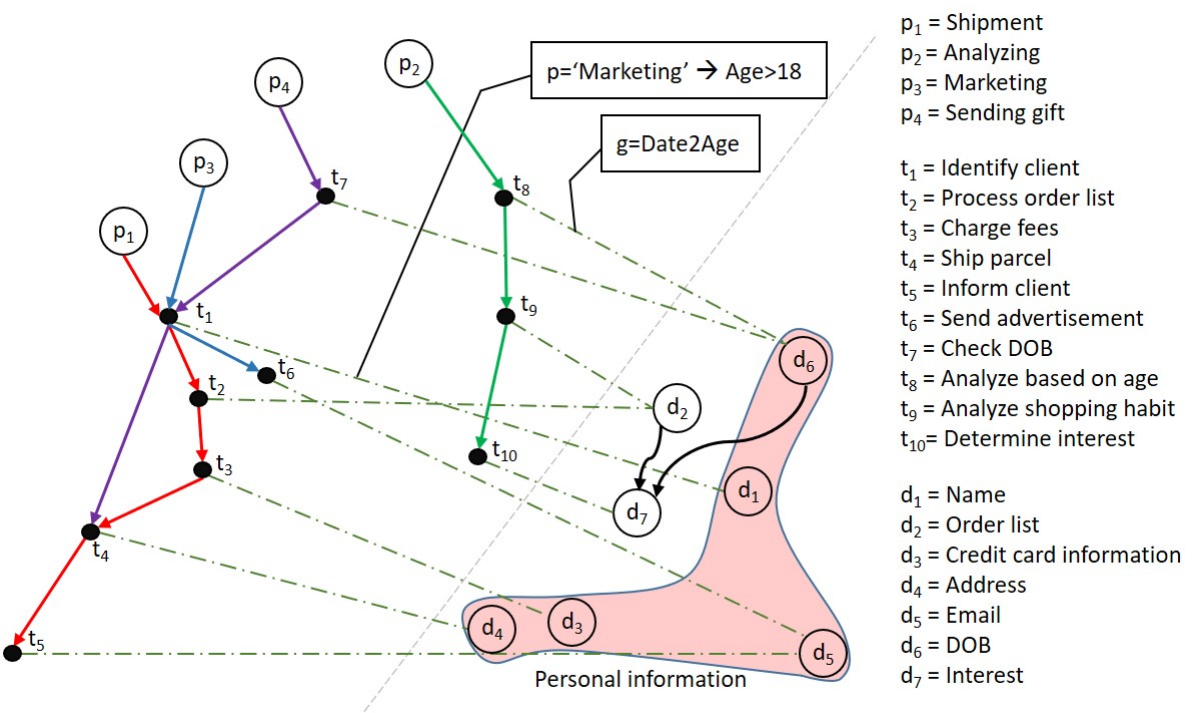}
	\caption{Purpose-attribute diagram}
	\label{fig:DiagramPurposeAttribute}
\end{figure}

Purpose-attribute heterogeneous permissions provide access to attributes for identified purposes. Since attributes are accessed by explicitly identified tasks, which are only accessed for defined purposes, the connection of purposes to attributes is via purposes' corresponding tasks. Figure \ref{fig:DiagramPurposeAttribute} illustrates these connections, listed in Table \ref{table:PurposeAttributePermissionInstances}, for our example privacy policy. For example, the task 'Identify client' ($t_1$) is connected to the 'Name' attribute ($d_1$), with the condition that, within the 'Marketing' purpose ($p_3$), the corresponding age must be over 18. An example of how an attribute can be connected to multiple purposes is illustrated with the 'Order list' attribute ($d_2$). The tasks 'Process order list' ($t_2$) and 'Analyze shopping habit' ($t_9$) connect to the attribute 'Order list' ($d_2$), so connections to this attribute exist for purposes 'Shipment' ($p_1$) and 'Analyzing' ($p_2$). For the 'Analyze based on Age', the 'DOB' attribute's value must be converted to age which is added to the corresponding connection in the diagram.

\clearpage
\subsubsection{Final Diagram for the Example Privacy Policy}
\begin{figure}[H]
 \centering
	\includegraphics[width=\textwidth]{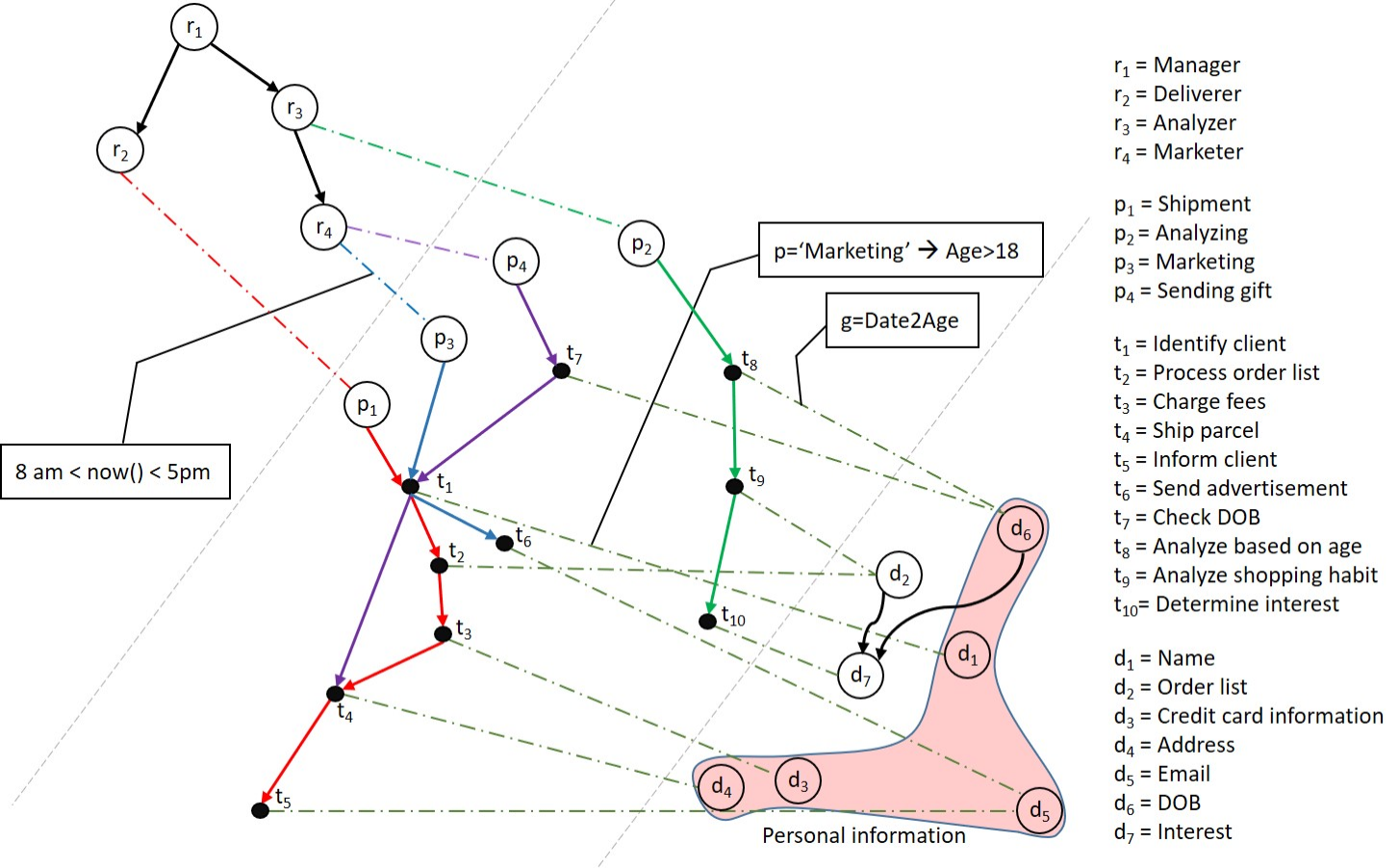}
	\caption{Final diagram for the example privacy policy}
	\label{fig:DiagramPrivacyPolicy}
\end{figure}

Figure \ref{fig:DiagramPrivacyPolicy} shows the complete PPPD, including components, and both homogeneous and heterogeneous connections identified in the example privacy policy in Appendix \ref{apx:ImgOnlShopPP}. 
\newline

Thus, to develop a PPPD we must first identify all of the components described in the privacy policy including roles, purposes, and data attributes (and attribute groups when applicable), and each is placed in their corresponding layer. 

Connections between the homogeneous component instances are identified to capture the structure in each layer. Operational features are defined by tasks that structure how purposes access data in the purpose layer. Finally, attribute aggregation and implicit attributes are captured in the attribute layer. These fully define all homogeneous connections for a particular policy. 

The final step is to capture the heterogeneous connections, including those between roles and purposes and between purposes and attributes. Ultimately, these connections represent permissions for roles to access purposes, and permissions for purposes to use attributes.

\clearpage
\section{Applying the PPPM}\label{sec:ApplyingPPPM}
By applying the PPPM, we develop a PPPD for the ChatterBaby\texttrademark\ application's privacy policy \cite{ChatterBaby2017}. This application, developed at UCLA \footnote{ChatterBaby\texttrademark's privacy policy exists as a pdf document so it is clearly intended to be a living document, which is completely appropriate for a such an application. Our assessment was undertaken based on its December 26$^{th}$, 2020 version and it is available at the following URL: \url{https://chatterbaby.org/files/view/download_files/Privacy_Policy_IRB.pdf} }, collects recordings of infants crying to interpret their needs. The collected audio files are also used to assess autism risk factors. We are not concerned about the utility, ethics, or capabilities of their application, but rather evaluate their privacy policy using our methodology. We generate the diagram and identify the privacy policy shortfalls by evaluating the result. 

\begin{figure}[H]
 \centering
	\includegraphics[width=0.9\textwidth]{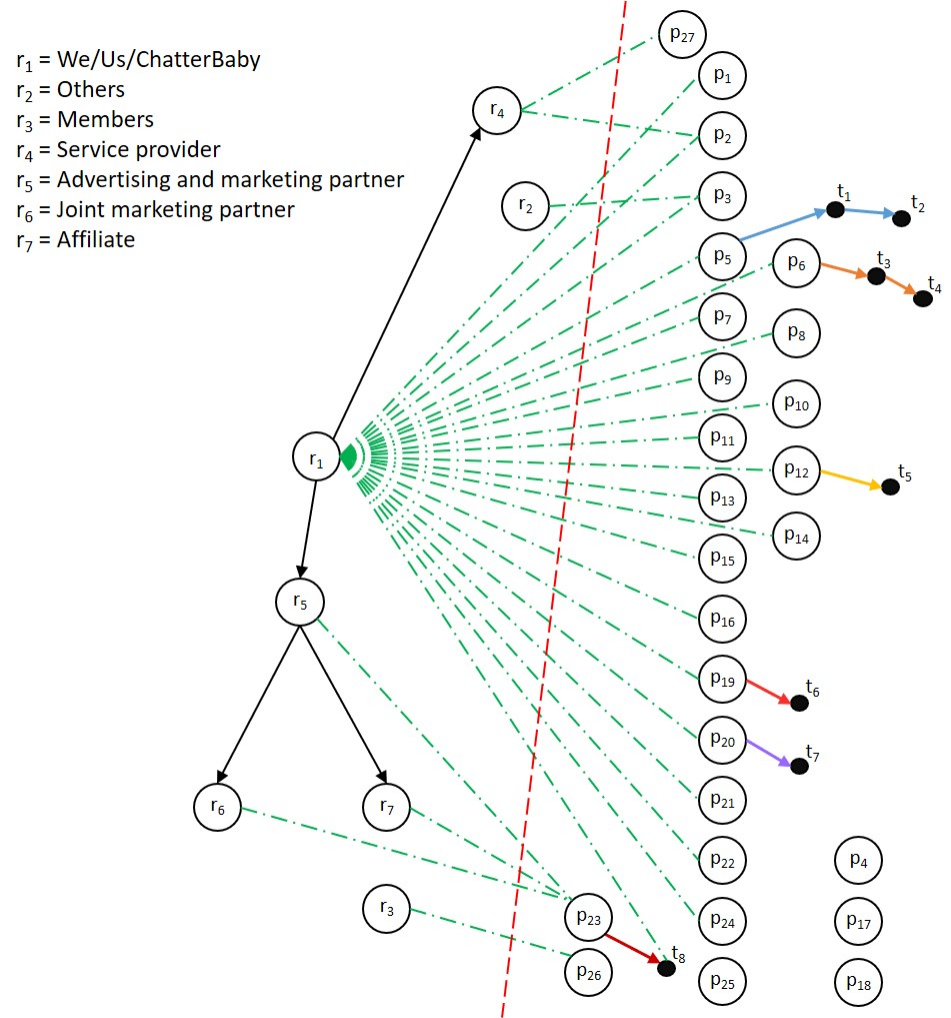}
	\caption{ChatterBaby\texttrademark\ role-purpose permission diagram}
	\label{fig:CBDiagramRolePurpose}
\end{figure}

Figure \ref{fig:CBDiagramRolePurpose} illustrates the role-purpose layer of the privacy policy. (All component instances and their connections identified from the privacy policy are listed in Appendix \ref{apx:CBCompNConn}). 

It is noteworthy that the role 'We/Us/ChatterBaby' has permission for many purposes, including purpose 'Any' ($p_{24}$). Although this approach of identifying the organization as a whole is often found in corporate privacy policies, it is problematic because no distinction can be made about which roles can actually access sensitive data because the policy explicitly allows anyone access within the organization. This single all-encompassing role, with universal permission, is an ideal illustration of how the PPPM approach highlights risky privacy policy features. As a counterpoint to this challenge, ChatterBaby\texttrademark\ also defines a role called 'Others' ($r_2$), which has permission to access the 'Identify' purpose ($p_3$). Unfortunately, this vague purpose is not defined, so the statement and its implications should be reviewed.

\begin{figure}[H]
 \centering
	\includegraphics[width=\textwidth]{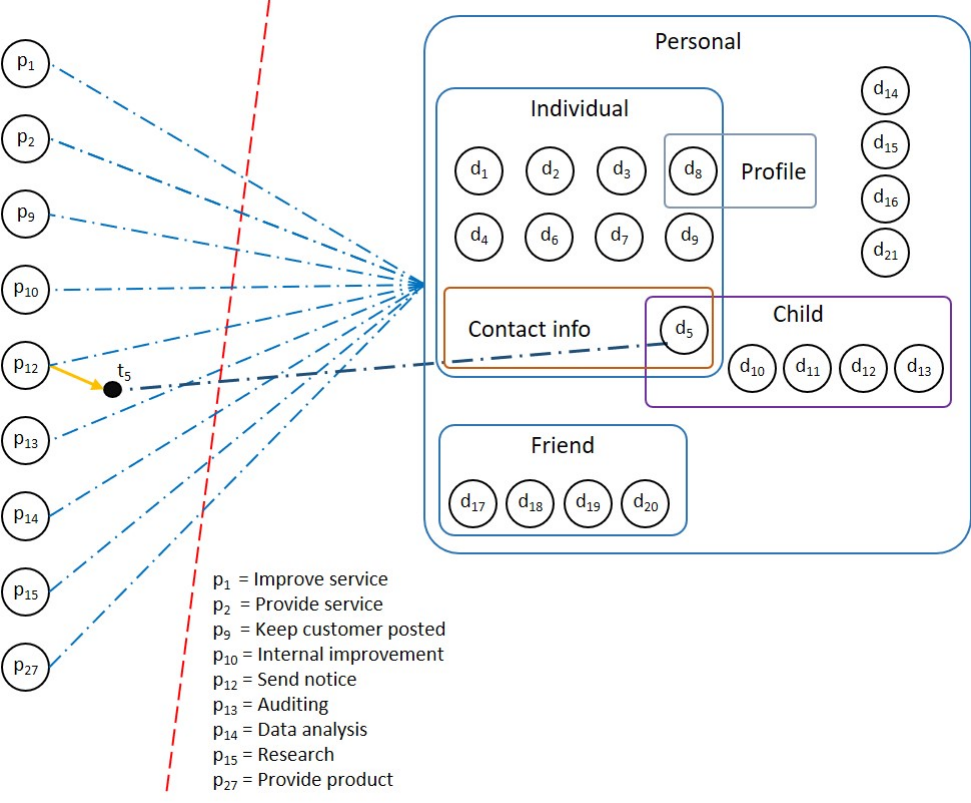}
	\caption{ChatterBaby\texttrademark \ heterogeneous permissions diagram to access the 'Personal' group}
	\label{fig:CBDiagramPurposeAttributePersonal}
\end{figure}

Figure \ref{fig:CBDiagramRolePurpose} highlights another key value of the PPPM. There is no role connected to 'Collect, measure, process autism risk' ($p_{4}$), 'Fighting spam/malware' ($p_{17}$), or 'Facilitate data collection' ($p_{18}$). This implies that any data attributes accessed for these purposes are not explicitly connected to a responsible role. This arises in the ChatterBaby\texttrademark \ policy because the relevant statements are written using a passive voice, so it is unclear who is responsible. While passive statements may not pose legal risks, and may even have advantages if challenged legally, they lack clarity about who accesses the data and for what purpose.
\newline
\clearpage
The complete list of all data attributes identified in the ChatterBaby\texttrademark \ privacy policy are found in Appendix \ref{apx:CBCompNConn}, and are organized into the corresponding attribute groups, as provided in Table \ref{table:CBAttributeGroup}. Note that the 'Individual' attribute group, which contains explicit personally identifying information, can be accessed for several purposes. In fact, the entire group of 'Personal' attributes contains what most would consider sensitive or personal data. 
\newline
Figure \ref{fig:CBDiagramPurposeAttributePersonal} depicts another part of ChatterBaby\texttrademark's PPPD, which highlights additional concerns. The policy statement: "From time to time, we may use your Personal Information to send important notices, such as communications about purchases and changes to our terms, conditions, and policies" provides purpose 'Send notice' ($p_{12}$) with access to all attributes in the 'Personal' group. On the other hand, the statement "if we believe that the changes are material, we'll let you know by [...] sending you an email or message about the changes" indicates that $p_{12}$ only has task 'Send email' ($t_{5}$), which accesses the 'Email' ($d_{5}$) attribute. Therefore, the permission of $p_{12}$ to access the personal group has no justification. In general, if a purpose is not connected to the attributes through tasks, then it is unclear how the data is used. This contradiction could be corrected by limiting the access available to purpose $p_{12}$.
\newline

\begin{figure}[H]
 \centering
	\includegraphics[width=0.9\textwidth]{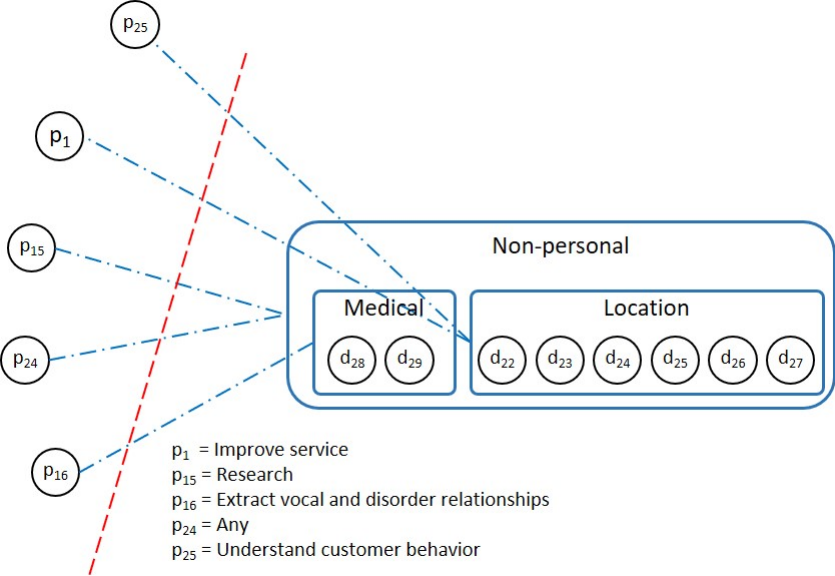}
	\caption{ChatterBaby\texttrademark \ heterogeneous permissions diagram to access the 'Non-personal' group}
	\label{fig:CBDiagramPurposeAttributeNon-Personal}
\end{figure}

Figure \ref{fig:CBDiagramPurposeAttributeNon-Personal} illustrates the data included in the 'non-personal' group. Although it is unclear how medical or location information can be considered 'non-personal' information, we will use this label, but note that the policy is likely implying that this data is anonymized sufficiently to be considered non-identifying. We wish to provide a PPPD reflective of the policy, but this terminology is misleading, and should be flagged through the PPPD definition process to highlight such narrative inconsistencies.
\newline

Setting the nomenclature aside for the moment, Figure \ref{fig:CBDiagramPurposeAttributeNon-Personal} highlights another concern in the policy. Purpose $p_{24}$, which represents 'Any' purpose has permission to use all the attributes in the 'Non-personal' group. This permission is explicit in the policy statement: "We may collect, use, transfer, and disclose non-personal information for any purpose." Given that medical and location information are subgroups of the 'Non-personal' attribute group, this would effectively allow for unrestrained access to this sensitive information. Our assumption that this data is anonymized is critical in such a situation, but it is based on a more tenuous assumption that the anonymization process is fully protective. It is also noteworthy that the use of universal access for 'Any' purpose is a clear violation of best practices in privacy protection and such a purpose should be flagged as a severe privacy risk.

\section{Discussion and Directions}
\subsection{PPPM Advantages}
Most of privacy policies today, do not specifically explain how the data is used or combined to generate additional information, and that is why they are vague. In our sample privacy policy, we clearly describe how data is combined to create new knowledge. For example, in the statement, “Our analyzers combine your date of birth, and shopping history to better understand your shopping habits, and predict your interests”, we specifically specified that DOB and shopping history are combined to find customer’s interests. This type of statement is not common in privacy policies, and we argue that this is a problem; therefore, we use PPPM to show this gap throughout the diagram. The power of the PPPM is its ability to highlight ambiguities and shortfalls in privacy policies, which can then be actioned by privacy officers, policy developers, and lawful organizations to correct policy shortfalls efficiently. We have provided a few examples of the shortfalls in Section \ref{sec:ApplyingPPPM}. The PPPD development process is structured to systematically identify and define privacy components. The component connections identify only explicit permissions associated with those components. This allows an organization to specify the components and their connections clearly and flag any inconsistent accesses once the system becomes operational. 

PPPM also provides an easy way to demonstrate an organization's practices and capture how client data is used. The resulting diagram could also help clients to understand the organization's privacy policy in a visual way.
PPPM is an abstract tool and does not require conforming to a rigorous definition of privacy or an abstraction of access. For example, earlier work required that the privacy context be defined, as illustrated by privacy conformance efforts in social networks, which first had to define classes of visibility such as 'Data provider', 'Friends', 'Friends of friends', 'Third parties', and 'All/world' \cite{barker2009data}. Unfortunately, the work on social networks using this categorization could not be ported to a different environment such as a hospital. The PPPM approach does not require the pre-definition of roles, purposes, or data attributes and types.

PPPM also specifies the connections between different components. Using some instances of components together in a privacy policy may result in impracticable policies or privacy violations. For example, any use of a universal access such as 'Any' purpose or 'All' data attributes is highlighted in the formation of the PPPD and can be carefully reviewed for potential (possibly unintentional) privacy risks. 

We believe that this modeling methodology would also allow for changes to privacy policies. As new statements are added or existing statements are adjusted, they could be evaluated using the existing PPPD, and more quickly assessed for nascent privacy risks. Developing a tool to allow for this evaluation and assessment to occur automatically is left as a future direction. 

\subsection{Limitations}
A clear limitation of the PPPM approach is that it only reviews how data is handled with respect to the privacy policy. If the policy does not disclose how data might be transferred to third-parties or used outside of the scope of the policy, there is no way to model potential privacy issues. This is not a fault of PPPM itself, but rather a pragmatic limitation of the policy. Understanding how to model the external flow of data beyond an organization's boundaries would be an interesting extension to the PPPM, so it is left as an opportunity for future research. 

PPPM itself is highly valuable but it relies on the organization to enforce the principles involved in the privacy policy. Another valuable direction would be to develop mechanisms that allow the PPPD to be used as input for an enforcement system to ensure that operational data only flows in accordance with the model. 

Using PPPM, the resulted diagram becomes more complex if the privacy policy is long. To address this issue, we are working on a user-friendly tool to give the user the ability to define "Zones" for different parts of the privacy policies. Nevertheless, the underneath concept of the tool will stay based on PPPM.

\subsection{Summary}
Often privacy policies lack a complete and clear explanation of how data is used. In many privacy policies, an organization's internal processes are explained in vague or ambiguous language and contain gaps and/or contradictions. PPPM can model privacy policies that allow for the identification of these shortfalls. A formally modeled privacy policy also allows for managing access permissions and information usage in an organization. This methodology is not domain-dependent, so when using PPPM, an organization’s database design does not need to be modified to accommodate the capturing of privacy policies. The database design maintains and expands on an ERD in that the PPPM creates a separate diagram representing the privacy policy. PPPM provides the privacy officers and policy designers with a diagram of their privacy policy. This diagram could enable the organization's clients to better understand the privacy policy that they may sign.
\newpage
% \bibliography{References.tex}

\clearpage
\appendixpage
\appendix
\renewcommand{\thesubsection}{Part \Roman{subsection}}

\section{ImaginaryOnlineShopping Privacy Policy} \label{apx:ImgOnlShopPP}
This privacy notice discloses the privacy practices for (ImaginaryOnlineShopping.com)\footnote{This privacy policy is a modified version of Eider House's privacy policy \cite{EiderHousePP}.}. This privacy notice applies solely to information collected by this website. It notifies you of the following:\\\\
1. What personally identifiable information is collected from you through the website, how it is used and with whom it may be shared.\\
2. Who uses your data in our company. \\
3. What choices are available to you regarding the use of your data.\\
4. The security procedures in place to protect the misuse of your information.\\
5. How you can correct any inaccuracies in the information.\\

\subsection{Information Collection, Use, and Sharing}\label{sec:InfoCollection}
We are the sole owners of the information collected on this site. We only have access to/collect information that you voluntarily provide via email or other direct contact from you. We will not sell or rent this information to anyone.
We will use your information to respond to you, regarding your purchases. We will not share your information with any third party outside of our organization, unless necessary to fulfill your request, e.g. to ship an order.

\par In order to use our website, you must first complete the registration form. During registration, you are required to provide certain personal information (such as name, email, and address). This information is used to contact you about the products/services on our site in which you have expressed interest. You may also provide personal demographic information (such as date of birth) about yourself, but it is not required.

\subsection{Access to your data}\label{sec:AccessYourData}
In our organization, only employees who need information to perform a specific job (for example, shipping, sending gift, or analyzing) are granted access to personal information.

\subsection{Orders and Shipment}\label{sec:OrderShipment}
We request additional information from you on our order form. To buy from us, you must provide contact information (like name and email address) and financial information (like credit card number). Deliverers ship orders that you place. To ship your orders, deliverers access your name, order list, credit card information, address, and email address to respectively identify you, process your order, charge fees, ship the parcel, and finally, inform you about the shipment. 

We use an outside shipping company to ship orders, and a credit card processing company to bill users for goods and services. These companies do not retain, share, store, or use personally identifiable information for any secondary purposes beyond filling your order.

\subsection{Marketing}\label{sec:Marketing}
From time-to-time, our analyzers perform analyses on your shopping history and date of birth to enhance our services. Our analyzers combine your date-of-birth and shopping history to better understand your shopping habits, and predict your interests; our marketers will then suggest products that might interest you. A marketer is an employee with a valid contract term, and they work under Analyzers' supervision; the manager supervises analyzers and deliverers. Marketing staff members will send you advertisements within business hours. To send advertisements, first marketers identify you by name, then use your email to send you suggestions and ads that might interest you. Our marketers will also send you a gift on your birthday. To send birthday gifts, marketers check your date of birth, identify you, and send a gift to your address. A customer's name can be used for marketing, if the customer is over 18 years old.

\subsection{Your Access to and Control Over Information:}\label{sec:YourAccess} 
Unless you ask us not to, we may contact you via email in the future to tell you about specials, new products or services, or changes to this privacy policy. You may opt out of any future contacts from us at any time. You can do the following at any time by contacting us via the email address or phone number given on our website:\\\\
•	See what data we have about you, if any.\\
•	Change/correct any data we have about you.\\
•	Have us delete any data we have about you.\\
•	Express any concern you have about our use of your data.\\

\subsection{Security} \label{sec:Security}
We take precautions to protect your information. When you submit sensitive information via the website, your information is protected both online and offline.
Wherever we collect sensitive information (such as credit card data), that information is encrypted and securely transmitted to us. You can verify this by looking for a lock icon in the address bar and looking for "https" at the beginning of the address of the Web page.
While we use encryption to protect sensitive information transmitted online, we also protect your information offline. The computers/servers in which we store personally identifiable information are kept in a secure environment. 

\clearpage
\section{ChatterBaby\texttrademark's Privacy Policy Components and Connection Lists} \label{apx:CBCompNConn}
\setcounter{table}{0}
\renewcommand{\thetable}{\ref{apx:CBCompNConn}\arabic{table}}

Table \ref{table:CBRoles} lists all of the roles mentioned in this privacy policy. Several statements in {ChatterBaby\texttrademark's} privacy policy introduce 'We', 'Us', and 'ChatterBaby\texttrademark' as a single role, for the first time, in the privacy policy. Although 'We/Us/ChatterBaby\texttrademark' seem like proper names, which refer to the same specific, distinct, legal entity, and choice-of-words is designed to be non-legalistic and friendly for the consumer, they do not precisely introduce the role in the company; rather, they introduce the role as the whole company. 
\newline
{
\setstretch{0.75}
\begin{table}[H]
    \centering
    \scalebox{0.85}
    {
    \begin{tabular}{c l}
        \toprule
        \textbf{Label} & \textbf{Role}\\
        \midrule
        r\textsubscript1 & We/Us/ChatterBaby \\
        r\textsubscript2 & Others \\
        r\textsubscript3 & Member \\
        r\textsubscript4 & Service provider \\
        r\textsubscript5 & Advertising and marketing partner\\
        r\textsubscript6 & Joint marketing partner \\
        r\textsubscript7 & Affiliate \\
        \bottomrule 
    \end{tabular} 
    }
    \caption{ChatterBaby\texttrademark's roles} \label{table:CBRoles} 
\end{table}
}

\vspace{10 mm}

Table \ref{table:CBPurposes} shows the list of all the purposes and their labels.

{\setstretch{0.75}
 \begin{table}[H]
 \centering
 \scalebox{0.85}
 {
 \begin{tabular}{c l l c l}
 \toprule
 \textbf{Label} & \textbf{Purpose} &&\textbf{Label} & \textbf{Purpose}\\
 \midrule
 p\textsubscript1 & Improve service && p\textsubscript{15} & Research \\
 p\textsubscript2 & Provide service && p\textsubscript{16} & Extract vocal and disorder relationships \\
 p\textsubscript3 & Identify && p\textsubscript{17} & Fighting spam/malware \\
 p\textsubscript4 & Collect, measure, \& process autism risk && p\textsubscript{18} & Facilitate data collection \\
 p\textsubscript5 & Extracting acoustic features && p\textsubscript{19} & Identify web browser \\ 
 p\textsubscript6 & Send alert && p\textsubscript{20} & Find site visit statistics \\ 
 p\textsubscript7 & Fulfill request && p\textsubscript{21} & Marketing \\
 p\textsubscript8 & Anti-fraud && p\textsubscript{22} & Send research participation request \\ 
 p\textsubscript9 & Keep customer posted && p\textsubscript{23} & Promote service \\
 p\textsubscript{10} & Internal improvement && p\textsubscript{24} & Any \\
 p\textsubscript{11} & Send service information && p\textsubscript{25} & Understand customer behavior \\
 p\textsubscript{12} & Send notice && p\textsubscript{26} & Facilitate interaction \\
 p\textsubscript{13} & Auditing && p\textsubscript{27} & Provide product \\
 p\textsubscript{14} & Data analysis \\ 
 \bottomrule 
 \end{tabular} 
 }
 \caption{ChatterBaby\texttrademark's purposes} \label{table:CBPurposes} 
\end{table}
}

\clearpage
Table \ref{table:CBAttributeGroup} shows the attribute list. The last column contains the attribute groups according to ChatterBaby\texttrademark's privacy policy. In this privacy policy, attributes are directly categorized into three main groups: 'Personal information', 'Non-personal information', and 'Other information'. Each group also has sub-groups. According to the privacy policy, 'Personal information' includes 'Individual', 'Friend', 'Child', 'Profile', and 'Contact information' groups. The 'Non-personal information' group includes 'Medical', and 'location' groups. Finally, the 'Other information' group includes 'Device-specific', and 'Browser' groups. In the statement "Where we use your data for direct marketing purposes...", the policy refers to accessing customers' data. Therefore, we consider 'Data' as a group that includes all the collected data about customers.
\newline
{\setstretch{0.75}
\begin{table}[H]
 \centering
 \scalebox{0.85}
 {
 \begin{tabular}{c l l}
 \toprule
 \textbf{Label} & \textbf{Attribute} & \textbf{Group} \\
 \midrule
d\textsubscript{1} & Name & Data, Personal, Individual \\
d\textsubscript{2} & Age & Data, Personal, Individual \\
d\textsubscript{3} & Mailing address & Data, Personal, Individual \\
d\textsubscript{4} & Phone number & Data, Personal, Individual \\ 
d\textsubscript{5} & Email address & Data, Personal, Individual, Child, Contact information \\ 
d\textsubscript{6} & Contact preferences & Data, Personal, Individual \\ 
d\textsubscript{7} & Credit card information* & Data, Personal, Individual \\
d\textsubscript{8} & Username & Data, Personal, Individual \\ 
d\textsubscript{9} & Password & Data, Personal, Individual \\ 
d\textsubscript{10} & Child's name & Data, Personal, Child \\ 
d\textsubscript{11} & Child's date of birth & Data, Personal, Child \\ 
d\textsubscript{12} & Week of delivery & Data, Personal, Child \\
d\textsubscript{13} & Child's gender & Data, Personal, Child \\
d\textsubscript{14} & Audio recording & Data, Personal \\ 
d\textsubscript{15} & Video data & Data, Personal \\ 
d\textsubscript{16} & Information from services & Data, Personal \\ 
d\textsubscript{17} & Friend's name & Data, Personal, Friend \\ 
d\textsubscript{18} & Friend's mailing address & Data, Personal, Friend \\ 
d\textsubscript{19} & Friend's email & Data, Personal, Friend \\ 
d\textsubscript{20} & Friend's phone number & Data, Personal, Friend \\
d\textsubscript{21} & Written contents & Data, Personal \\
d\textsubscript{22} & Language & Data, Non-Personal, Location \\
d\textsubscript{23} & Zip code & Data, Non-Personal, Location \\ 
d\textsubscript{24} & Area code & Data, Non-Personal, Location \\ 
d\textsubscript{25} & Referrer URL & Data, Non-Personal, Location \\ 
d\textsubscript{26} & Location & Data, Non-Personal, Location \\
d\textsubscript{27} & Time zone & Data, Non-Personal, Location \\
d\textsubscript{28} & Medical history & Data, Non-Personal, Medical \\
d\textsubscript{29} & Autism risk factors & Data, Non-Personal, Medical \\
d\textsubscript{30} & IP address & Data, Other, Browser \\ 
d\textsubscript{31} & Cookies & Data, Other, Browser \\ 
d\textsubscript{32} & Device identifier & Data, Other \\
d\textsubscript{33} & Network information & Data, Other \\
d\textsubscript{34} & Hardware model & Data, Other \\ 
d\textsubscript{35} & Device interaction & Data, Other \\
 \bottomrule 
 \end{tabular} 
 }
 \caption{ChatterBaby\texttrademark's attributes}
 \label{table:CBAttributeGroup} 
\end{table}
}

* ChatterBaby\texttrademark's privacy policy includes two statements regarding collecting billing information. The statement "we may collect a variety of information, including your name, age, mailing address, phone number, email address, contact preferences, credit card information, username and password" specifies that 'Credit card information' might be collected. Alternatively, the statement "We will not collect billing information, as our service is free" specifies that no billing information is collected. Nevertheless, we include the 'Credit card information' in our diagram. This contradiction is highlighted though the process of extracting the information.
\vspace{10 mm}

Table \ref{table:CBRoleStructure} shows all of the roles' connections in the privacy policy. These connections are used to form a structure for the role layer. 

{\setstretch{0.75}
\begin{table}[H]
 \centering
 \scalebox{0.85}
 {
 \begin{tabular}{l l}
 \toprule
 \textbf{Superior} & \textbf{Inferior} \\
 \midrule
 We & Service Provider \\
 We & Advertising and marketing partner \\ 
 Advertising and marketing partners & Joint marketing partner \\
 Advertising and marketing partner & Affiliate \\
 \bottomrule 
 \end{tabular} 
 }
 \caption{ChatterBaby\texttrademark's role connections} 
 \label{table:CBRoleStructure}
\end{table}
}

\vspace{10 mm}

Purposes and their tasks are listed 
in Table \ref{table:CBPurposeStructure} if they are specified in the privacy policy.

{\setstretch{0.75}
\begin{table}[H]
 \centering
 \scalebox{0.85}
 {
 \begin{tabular}{l l}
 \toprule
 \textbf{Purpose} & \textbf{Tasks} \\
 \midrule
 Extracting acoustic features & Process audio recordings, process video data \\
 Send alert & Process service information, Email alert \\
 Send notice & Email notice \\
 Identify web browser & Identify web browser \\
 Find site visit statistic & Find site visit statistic \\ 
 Promote service & Disclose information \\
 \bottomrule 
 \end{tabular} 
 }
 \caption{ChatterBaby\texttrademark's purpose connections} 
 \label{table:CBPurposeStructure}
\end{table}
}
\vspace{10 mm}

In ChatterBaby\texttrademark's privacy policy, no statements provide accurate information about whe-ther any attributes are combined, or what new attributes are created. Therefore, we are not able to establish any connections between the attributes.

\clearpage
The privacy policy contains statements that specify permissions for roles to use purposes. Table \ref{table:CBRolePurpose} shows these permissions.

{\setstretch{0.75}
\begin{table}[H]
 \centering
 \scalebox{0.85}
 {
 \begin{tabular}{l l l}
 \toprule
 \textbf{Role} & \textbf{Purpose} & \textbf{Condition} \\
 \midrule
 We & Provide service & \\ 
 We & Improve service & \\
 Others & Identify & \\
 We & Identify & \\
 We & Extracting acoustic features & \\
 We & Send alert & \\
 ChatterBaby(We) & Fulfill request & \\
 ChatterBaby(We) & Anti-fraud & \\
 We & Keep customer posted & \\
 We & Internal improvement & \\
 We & Send service information & \\
 We & Send notice & \\
 We & Auditing & \\
 We & Data analysis & \\
 We & Research & \\ 
 We & Any & \\
 We & Understand customer behavior & \\
 We & Extract vocal and disorder relationships & \\
 We & Identify web browser & \\ 
 We & Find site visit statistics & \\ 
 We & Marketing & \\ 
 We & Send research participation request & \\
 Member & Facilitate interaction & \\
 Service provider & Provide service & \\
 Service provider & Provide product & \\
 Advertising and marketing partner & Promote service & \\
 Joint marketing partner & Promote service & \\
 Affiliate & Promote service & \\
 \bottomrule 
 \end{tabular} 
 }
 \caption{ChatterBaby\texttrademark's role-purpose permissions} \label{table:CBRolePurpose}
\end{table}
}
\clearpage
Table \ref{table:CBPurposeAttribute} lists purposes' permissions for accessing attribute groups, and \ref{table:CBTaskAttribute} lists the tasks and their required attributes.

{\setstretch{0.75}
\begin{table}[H]
 \centering
 \scalebox{0.85}
 {
 \begin{tabular}{l l l}
 \toprule
 \textbf{Purpose} & \textbf{Attribute group} & \textbf{Condition} \\
 \midrule
 Improve service & Personal, Location & \\
 Provide service & Personal & \\
 Identify & Individual & \\
 Collect, measure, and process autism risk & Child & \\
 Extracting acoustic features & Acoustic & \\
 Provide service & Friend & \\ 
 Fulfill request & Friend & \\
 Anti-fraud & Friend & \\ 
 Keep customer posted & Personal & \\
 Internal improvement & Personal & \\
 Send service information & Contact information & Consent = True \\
 Send notice & Personal & \\
 Auditing & Personal & \\
 Data analysis & Personal & \\
 Research & Personal & \\
 Research & Non-personal & \\ 
 Understand customer behavior & Location & \\ 
 Extract vocal and disorder relationships & Medical & \\
 Fighting spam/malware & Browser &\\ 
 Facilitate data collection & Browser &\\ 
 Identify web browser & Cookies & \\
 Find site visit statistics & Cookies & \\ 
 Marketing & Data & Subscription = True\\
 Send research participation request & Contact information & \\
 Facilitate interaction & Profile\\
 Provide product & Personal & \\
 Promote services & Service information & \\
 \bottomrule 
 \end{tabular} 
 }
 \caption{ChatterBaby\texttrademark's purpose-attribute permissions} \label{table:CBPurposeAttribute}
\end{table}
}

{\setstretch{0.75}
\begin{table}[H]
 \centering
 \scalebox{0.85}
 {
 \begin{tabular}{l l l l l}
 \toprule
 \textbf{Label} & \textbf{Task} & \textbf{Attribute} & \textbf{Condition} & \textbf{Granularity} \\
 \midrule
 t\textsubscript{1} & Process audio recording & Audio recording & \\
 t\textsubscript{2} & Process video data & Video recording & \\
 t\textsubscript{3} & Process info from services & Information from services & \\
 t\textsubscript{4} & Send alert & Email & \\
 t\textsubscript{5} & Send notice & Email & \\
 t\textsubscript{6} & Identify web browser & Cookies & \\
 t\textsubscript{7} & Find site visit statistics & Cookies & \\ 
 t\textsubscript{8} & Disclose information & Service information & \\
 \bottomrule 
 \end{tabular} 
 }
 \caption{ChatterBaby\texttrademark's Purpose-attribute (tasks) permissions} \label{table:CBTaskAttribute}
\end{table}
}

\clearpage
\section{ChatterBaby\texttrademark's Privacy Policy Permission Diagram} \label{apx:CBPPPD}
By using the information in Table \ref{table:CBRoles}, we create a node for each role in the privacy policy. We then use Table \ref{table:CBRoleStructure} to add the connections between the roles. Diagram \ref{fig:CBDiagramRole} depicts the complete role layer. 

\setcounter{figure}{0}
\renewcommand{\thefigure}{\ref{apx:CBPPPD}\arabic{figure}}

\begin{figure}[H]
 \centering
	\includegraphics[width=0.65\textwidth]{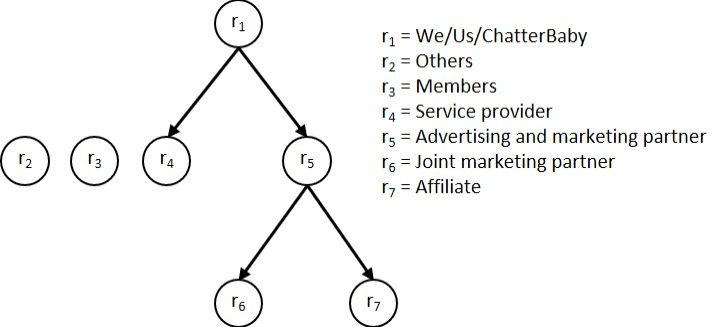}
	\caption{ChatterBaby\texttrademark's role structure diagram}
	\label{fig:CBDiagramRole}
\end{figure}

\begin{figure}[H]
 \centering
	\includegraphics[width=\textwidth]{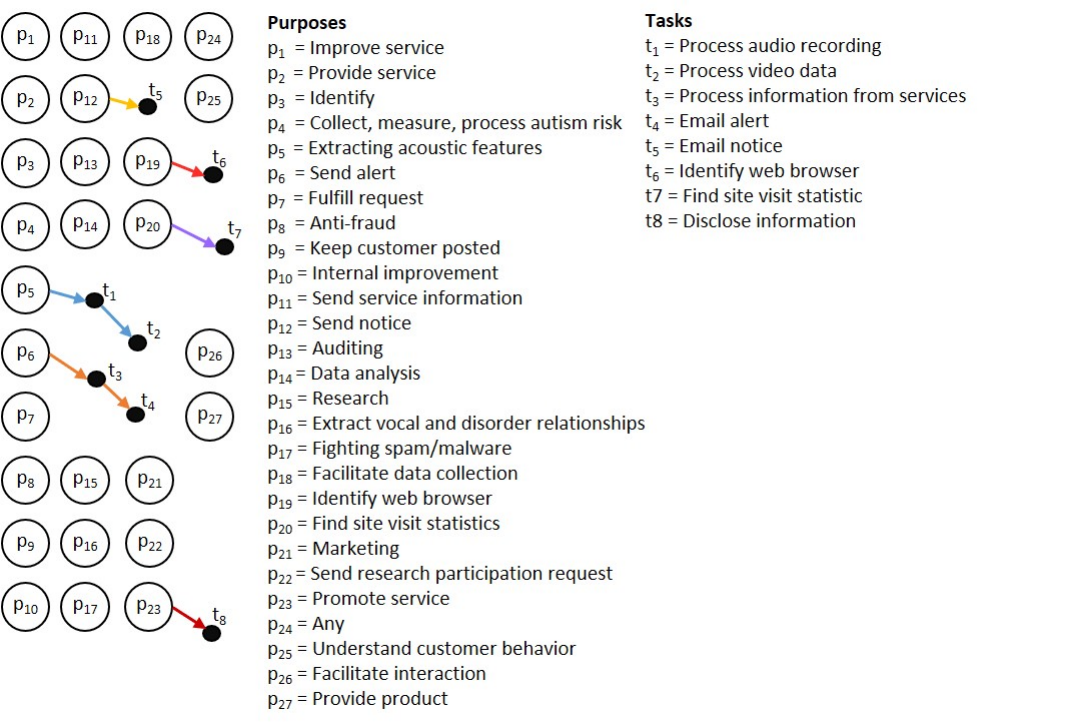}
	\caption{ChatterBaby\texttrademark's purpose structure diagram}
	\label{fig:CBDiagramPurpose}
\end{figure}

The ChatterBaby\texttrademark\ privacy policy diagram's purpose layer is created by adding a node for each purpose listed in Table \ref{table:CBPurposes}. If a purpose's tasks are provided, we illustrate them in the diagram, using the information in Table \ref{table:CBPurposeStructure}. Figure \ref{fig:CBDiagramPurpose} depicts the complete purpose layer.
\newline

\begin{figure}[H]
 \centering
	\includegraphics[width=\textwidth]{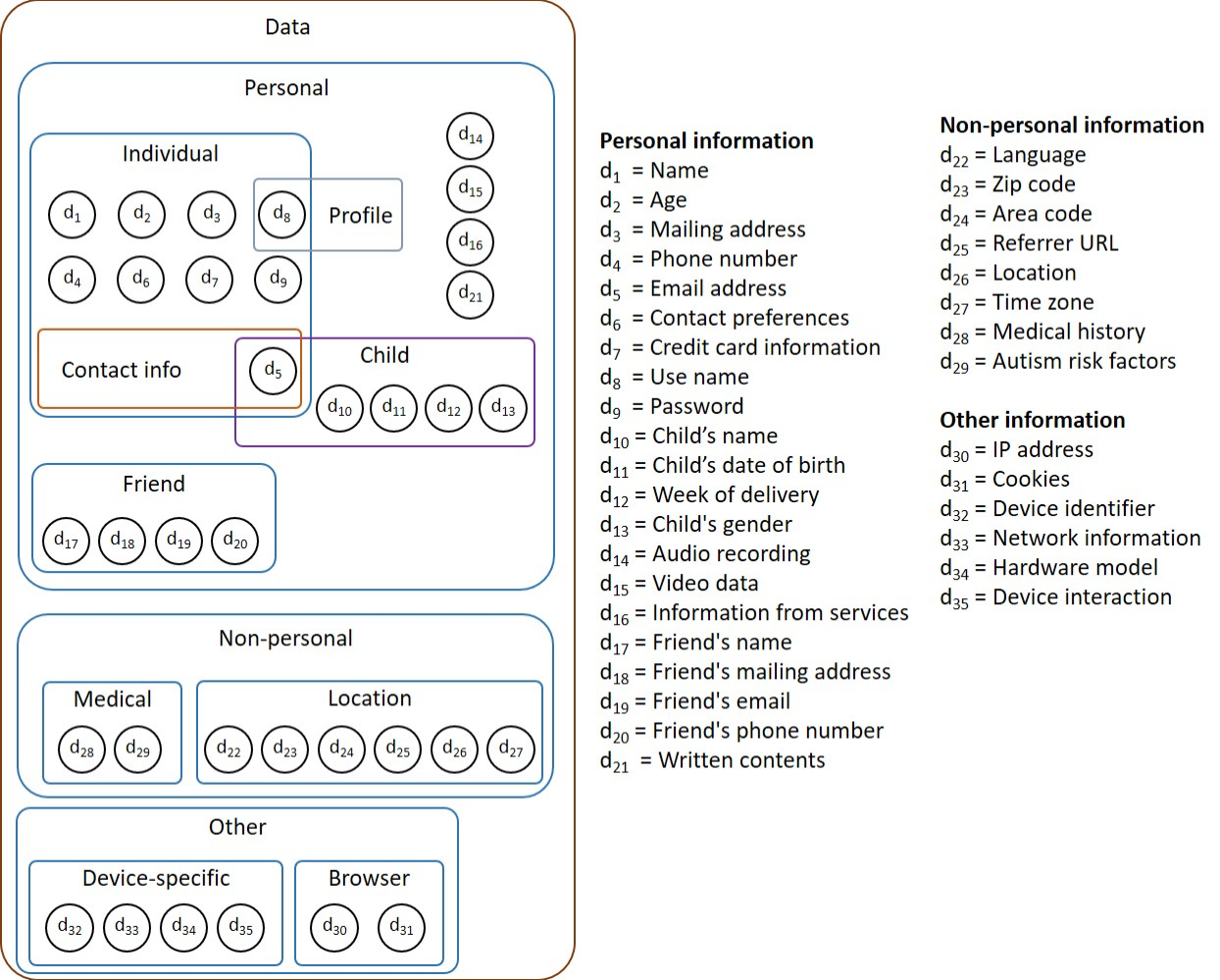}
	\caption{ChatterBaby\texttrademark's attribute structure diagram}
	\label{fig:CBDiagramAttribute}
\end{figure}

Figure \ref{fig:CBDiagramAttribute} shows the attribute layer in which attributes are categorized into their groups using information in Table \ref{table:CBAttributeGroup}. Since the privacy policy does not explicitly specify the attributes' connections, there are no edges in this layer. 
\newline

\begin{figure}[H]
 \centering
	\includegraphics[width=\textwidth]{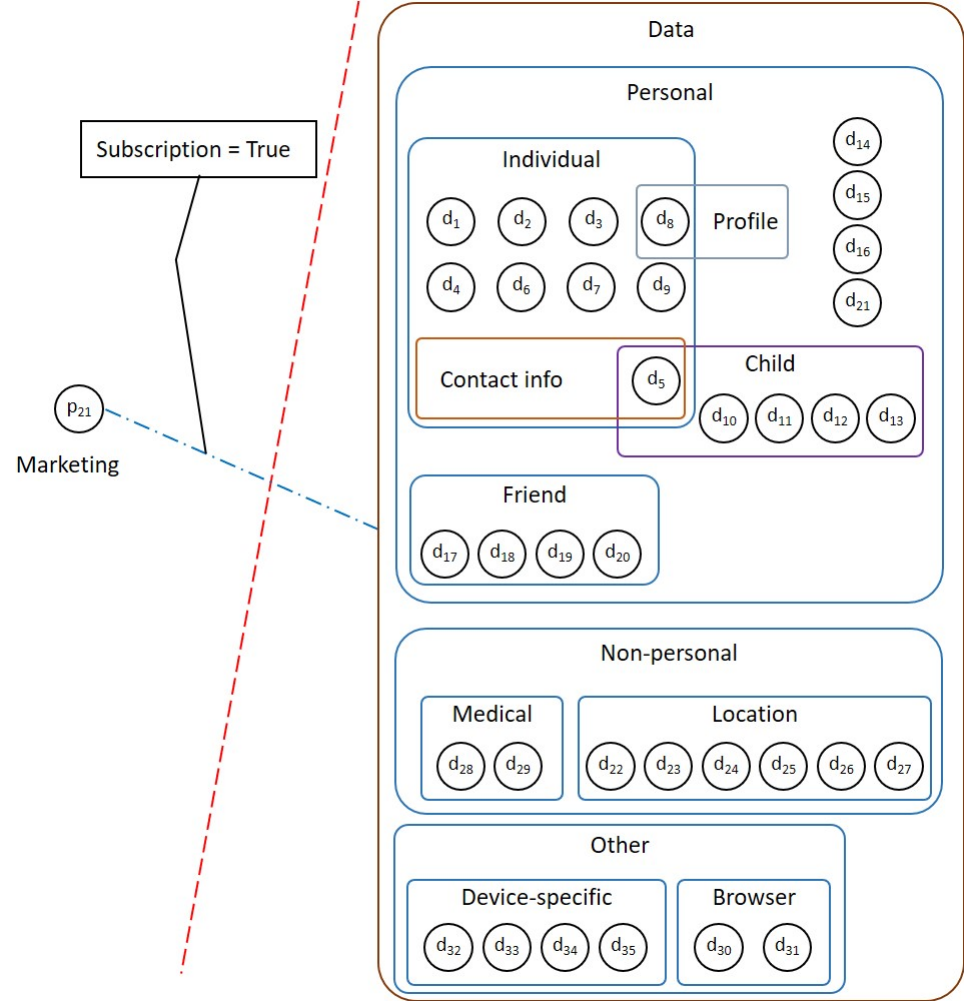}
	\caption{ChatterBaby\texttrademark's heterogeneous permissions to access the 'Data' group}
	\label{fig:CBDiagramPurposeAttributeData}
\end{figure}

\clearpage
The diagram in Figure \ref{fig:CBDiagramPurposeAttributeData} shows that if the customer subscribes, all their data can be used for the 'Marketing' purpose. This permission is a result of the statement "Where we use your data for direct marketing purposes, you can always object using the unsubscribe link in such communications or changing your account settings." This statement has 'We' as a role, 'Marketing' as a purpose, and 'Data' as a group that includes all attributes. 
\newline

\begin{figure}[H]
 \centering
	\includegraphics[width=\textwidth]{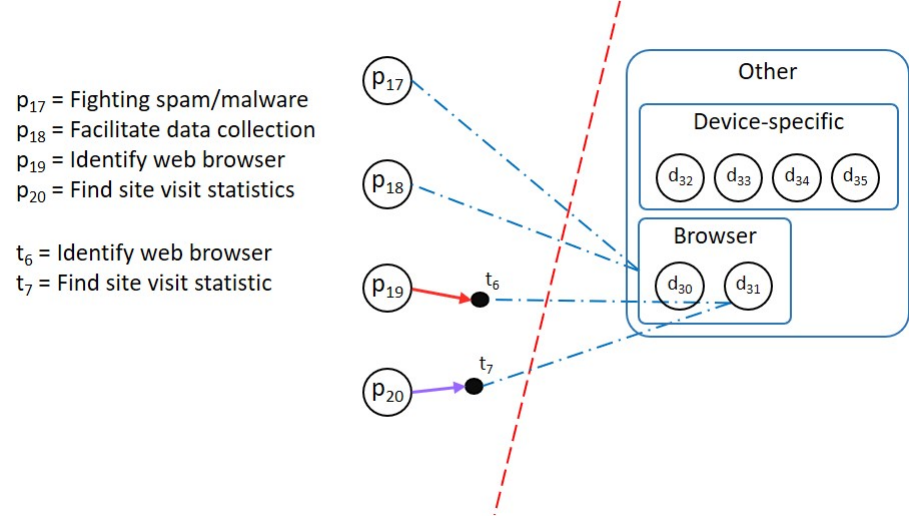}
	\caption{ChatterBaby\texttrademark's heterogeneous permissions to access the 'Other' group}
	\label{fig:CBDiagramPurposeAttributeOther}
\end{figure}

\clearpage
Figures \ref{fig:CBDiagramPurposeAttributeOther} shows permissions for the 'Other' group and its subgroups, capturing the statement "We may collect some device-specific information if you access the Services using a mobile device. Device information may include but is not limited to unique device identifiers, network information, and hardware model, as well as information about how the device interacts with our Services." While the data items of the attributes in the 'Device-specific' group may be collected, no statement in the privacy policy specifies a purpose for collecting them. 
This shortfall is illustrated in Figure \ref{fig:CBDiagramPurposeAttributeOther}, where the attributes in the 'Device-specific' group are not connected to any purpose.

\end{document}